\documentclass[12pt]{article}

\begin{document}

\begin{center}
{\LARGE WINHAC -- the Monte Carlo event generator for single  \\\vspace{1mm}
       $W$-boson production in hadronic collisions$^{\star}$}

\end{center}

\vspace{1mm}
\begin{center}

{\Large\bf Wies{\l}aw P{\l}aczek}\\
\vspace{4mm}
        Marian Smoluchowski Institute of Physics, Jagiellonian University,\\
        ul.\ Reymonta 4, 30-059 Krak\'ow, Poland.\\
\vspace{1mm}
        E-mail: {\sf Wieslaw.Placzek@uj.edu.pl}
\end{center}

\vspace{15mm}

\begin{abstract}
\noindent
We present the Monte Carlo event generator WINHAC for the 
charged-current Drell--Yan process, i.e. single $W$-boson production 
 with leptonic decays in hadronic collisions. It features multiphoton radiation
within the Yennie--Frautschi--Suura exclusive exponentiation scheme with
${\cal O}(\alpha)$ electroweak corrections. For the initial-state QCD/QED parton
shower and hadronisation it is interfaced directly to PYTHIA. 
It implements the options for proton--proton, proton--antipronton and any 
nucleus--nucleus collisions. 
Moreover, it includes the polarized $W$-boson production and the neutral 
current Drell--Yan process, both at the Born level. Generation of weighted as
well as unweighted (weight$=1$) events is possible. WINHAC has been thoroughly
tested numerically and cross-checked with independent Monte Carlo programs, 
such as HORACE and SANC; in the so-called tuned comparisons, an agreement 
at the sub-per-mil level has been reached. 
It has also been used as a basic tool for developing and testing some new
measurement methods of the Standard Model parameters at the LHC.
\end{abstract}

\vspace{15mm}

\begin{center}
{\em Talk given at the European Physical Society Europhysics Conference on High Energy Physics,
EPS-HEP 20009, July 16--22 Krakow, Poland.}
\end{center}

\vspace{15mm}

\footnoterule
\noindent
{\footnotesize \noindent
\begin{itemize}
\item[${}^{\star}$]
The work is partly supported by 
the program of the French--Polish cooperation between IN2P3 and COPIN 
No.\ 05-116 
and 
by the EU Marie Curie Research Training Network grant 
        under the contract No.\ MRTN-CT-2006-035505.
\end{itemize}
}

\clearpage

\section{Introduction}

Motivations for investigating the Drell--Yan processes, in particular
that involving the $W$-bosons, at the Tevatron and LHC can be found e.g.\ in 
Refs.~\cite{Baur:1998kt,Haywood:1999qg,Krasny:2005cb,Alcaraz:2006mx,Krasny:2007cy,Gerber:2007xk,Fayette:2008wt,Fayette:2009zi,Upcoming_MW,Buttar:2008jx}.

In theoretical descriptions of the Drell--Yan processes reaching a
sufficiently high precision for the Tevatron and LHC experiments 
requires including besides the QCD effects also the electroweak
(EW) corrections, in particular the QED final-state radiation (FSR) 
\cite{Baur:1998kt}. It was shown in Ref.~\cite{CarloniCalame:2003ux} 
that including 
${\cal O}(\alpha)$ EW corrections may be not sufficient because 
the higher-order FSR effects can shift $M_W$ by $\sim 10\,$MeV.
    
Regions of a high transverse momentum of a charged lepton or a large $W$-boson
transverse mass will be used at the LHC for various ``new physics'' searches.
In such regions the ${\cal O}(\alpha)$ EW corrections beyond the pure FSR 
can reach the size of $20$--$30\%$ \cite{Dittmaier:2001ay,Baur:2004ig}. 
Given high 
statistics of $W$-boson data expected at the LHC, these effects need to be 
included in order to have precise enough description of the Standard Model
(SM) background. 

Drell--Yan processes will also be treated at the LHC as the so-called 
`standard candle' processes (detector calibration, normalisation, etc.).
Therefore, their precise theoretical description is very important. 
In order to be fully exploited by the experiments, such a description 
ought to be provided in the form of a Monte Carlo event generator.  
Several semi-analytical programs as well as Monte Carlo integrators and 
event generators for Drell--Yan processes including QCD and/or EW effects 
have been developed in recent years, 
see e.g.\ Refs.~\cite{Gerber:2007xk,Buttar:2008jx}. 

In this paper we briefly describe the Monte Carlo event generator 
WINHAC~\cite{WINHAC:MC} dedicated to the charged-current Drell--Yan process.
Section~2 is devoted to the QED FSR, while in Section~3 we describe inclusion
of the ${\cal O}(\alpha)$ EW corrections. Finally, 
Section~4 contains summary and outlook.

\section{QED FSR}


Our
first step in the development of WINHAC was to take into account
effects of multiphoton FSR \cite{Placzek2003zg}, based on the 
Yennie--Frautschi--Suura (YFS) exclusive exponentiation~\cite{yfs:1961}.
The total cross section for a parton-level process 
\begin{equation}
q_1(p_1) + \bar{q}_2(p_2) \longrightarrow W^{\pm}(Q) 
 \longrightarrow l(q_l) + \nu(q_{\nu})
              + \gamma(k_1) + \ldots + \gamma(k_n),
 \:\:\: (n = 0,1,\ldots)
\label{eq:multiFSR}
\end{equation}
can be expressed as
\begin{equation}
\sigma^{tot}_{\rm YFS} = \sum_{n=0}^{\infty} \int \frac{d^3q_l}{q_l^0}
  \frac{d^3q_{\nu}}{q_{\nu}^0}\rho_n^{(1)}(p_1,p_2,q_1,q_2,k_1,\ldots,k_n),
\label{eq:sigtot}
\end{equation}
where
\begin{eqnarray}
\rho_n^{(1)}  & = &  \, e^{Y(Q,q_l;k_s)}\,\frac{1}{n!}\, 
  \prod_{i=1}^n \frac{d^3k_i}{k_i^0} \tilde{S}(Q,q_l,k_i)\theta(k_i^0-k_s)
  \,\delta^{(4)}\left(p_1 + p_2 - q_l - q_{\nu} - \sum_{i=1}^n k_i\right) 
  \nonumber \\ 
  &\times &
  \left[\,\bar{\beta}_0^{(1)}(p_1,p_2,q_l,q_{\nu}) + \sum_{i=1}^n 
   \frac{\bar{\beta}_1^{(1)}(p_1,p_2,q_l,q_{\nu},k_i)}{\tilde{S}(Q,q_l,k_i)}\,
  \right].
\label{eq:rhon}
\end{eqnarray}
The exponent in the above equation is the so-called YFS form factor in which
all the infrared (IR) singularities coming from virtual and real photons 
are re-summed (in a gauge-invariant way) to the infinite order and properly 
cancelled.
The rest of the equation is the accordingly reorganized perturbative series, 
which in the current version of WINHAC is truncated at ${\cal O}(\alpha)$. 
In particular, the $\bar{\beta}_0^{(1)}$ and $\bar{\beta}_1^{(1)}$ functions 
are the so-called non-IR YFS residuals corresponding to zero and one real
hard-photon emission, respectively. 
This means that the former contains pure virtual
corrections while the latter the pure real-photon ones. 
All details are given in Ref.~\cite{Placzek2003zg}.

Based on the above formula an appropriate Monte Carlo (MC) algorithm was
constructed. We employed the importance sampling method by simplifying
step-by-step the starting formula~(\ref{eq:sigtot}) until the simple enough
probability density
was obtained that could be generated with the use of basic MC methods. All these
simplifications were compensated with corresponding MC weights. From these
MC weights the full event weight was constructed. It can be used directly
in weighted-event mode, e.g. for histogramming, or it can be applied 
in a rejection loop to construct weight$=1$ events in unweighted-event mode. 
In the weighted-event mode the program calculates for each event a large 
array of the MC weights corresponding to various effects, corrections, etc.   
The parton level cross section is convoluted with a product of 
two parton distribution functions (PDFs) corresponding to two colliding
beams: proton--proton, proton--antiproton or any nucleus--nucleus.
Values of Bjorken-$x$'s of a quark and an antiquark participating in the hard
process are generated from a $2$-dimensional distribution, given by the product
of two PDFs convoluted with a simplified (``crude'') parton-level cross section,
with the help of the adaptive cellular MC sampler FOAM~\cite{Foam:2003}.
Then, the quark and antiquark flavours are randomly chosen according to their
relative contributions to the cross section.

After constructing the Monte Carlo event generator WINHAC we performed several
numerical tests, both internal and external. 
The important external test was the comprehensive comparison with
the independent MC program HORACE~\cite{CarloniCalame:2003ux} which includes
higher-order FSR corrections through a QED parton
shower algorithm. We found the agreement between the two programs at the
per-mil level, see Ref.~\cite{CarloniCalame:2004qw} for details. 
This is very important since the two programs are based on very different 
approaches to the QED corrections. 
It suggests that major parts of the higher-order FSR corrections are included 
in both programs.    

\section{EW corrections}

Strictly speaking, for the charged-current electroweak processes 
it is not possible
to split the EW corrections into the pure QED and pure weak ones
in a gauge-invariant way.
However, one can find gauge-invariant subsets of QED corrections and
separate them from the rest. Such a separation is in general not unique,
as one can identify different subsets of gauge invariant QED corrections.
In our definition of the QED FSR corrections described the previous section
we used the prescription of Marciano and Sirlin (M\&S) given in 
Ref.~\cite{Marciano:1973}. Then we included the rest of the electroweak
corrections for the FSR in the pole approximation (PA). 
These corrections have been
provided to us in the form of the Fortran-code module by the SANC group
\cite{Arbuzov:2005dd}. Numerically they are small -- at the level of 
$0.2$--$0.3\%$ -- and almost constant, so that they are not important for 
the $W$-mass measurements.


The main problem that appears in including the full ${\cal O}(\alpha)$ EW 
corrections (beyond PA) to the charged-current Drell--Yan process concerns
the QED initial-state radiation (ISR), i.e. photon radiation from incoming
quarks. In EW calculations the collinear ISR singularities are usually 
regularized with finite quark masses $m_q$. As a results, terms with
$\ln(\hat{s}/m_q^2)$ appear in theoretical predictions.  
However, $m_q$ in this context is not a well-defined quantity, mainly because
of the QCD effects. 
In fact, these singularities can be absorbed into the PDFs in a similar way
as the ones in QCD, which is usually done in the $\overline{\rm MS}$ or DIS
scheme, see e.g.\ Refs.~\cite{Dittmaier:2001ay,Baur:2004ig,Arbuzov:2005dd}.
QCD parton-shower generators, such as PYTHIA or HERWIG, usually generate also
the QED ISR, however they do it in neither of the above schemes.
So, combining such $\overline{\rm MS}$ or DIS-subtracted EW corrections
with parton shower generators would result in some mismatch. 
Besides, the above schemes are not well suited for fully exclusive Monte Carlo
generators, as they are defined for partly integrated quantities.

Therefore, we take a different approach to the QED ISR.
We identify the QED gauge-invariant subsets of the ISR corrections 
and subtract them from the total ${\cal O}(\alpha)$ EW correction,
leaving responsibility for their computation to parton-shower generators. 
In our opinion, it is better if the parton-shower generator treats both
the QCD and QED ISR since the QED radiation is strongly affected by the
QCD one. Of course, the parton-shower programs usually include the QED ISR
in some approximation. However, the QED ISR corrections are small and flat
for the main $W$-boson observables~\cite{Baur:1998kt}. 
Moreover, at the LHC the $W$-boson distributions will have to be normalized
to the corresponding $Z$-boson ones in order to avoid some large uncertainties,
e.g.\ coming from the PDFs parametrisations, $W$-boson $p_T$-spectrum 
predictions, lepton energy scale determination, etc. 
In such ratios most of the QED ISR uncertainties will also cancel. 

In WINHAC we have implemented three schemes for ISR subtraction: the one
similar to the M\&S prescription for the FSR, 
the YFS-motivated scheme and the one
from Ref.~\cite{Wackeroth:1996}. In fact, in the latter two schemes the ISR
virtual plus soft real-photon corrections consist of the YFS form factor plus 
the terms $\sim (1/2)Q_i^2[\ln(s/m_i^2) - 1]$, however our own formulae
differ from those of Ref.~\cite{Wackeroth:1996} by some constant terms.
Similar schemes can also be applied to the FSR and initial-final 
interferences -- resulting in a separation of the 
${\cal O}(\alpha)$ EW corrections into the `QED-like' corrections and
the `weak-like' corrections. Such a splitting is, of course, not unique
and it differs in various schemes by some constant terms.
Our numerical tests show that the YFS-motivated scheme for the ISR 
is very close to the $\overline{\rm MS}$ scheme 
-- the relative difference is $<0.05\%$.

Similarly as in the case of the FSR, for the full ${\cal O}(\alpha)$ EW 
virtual corrections we use the Fortran-code modules from the 
SANC library \cite{Arbuzov:2005dd}. 
Both virtual and real-photon corrections are implemented on top of the
formula~(\ref{eq:sigtot}) through appropriate MC weights.
In order to test their implementation,
we performed numerical comparisons at ${\cal O}(\alpha)$ of WINHAC and the 
SANC MC integrator based on VEGAS. We have found a very good agreement
-- at the sub-per-mil level -- between two programs for the total cross
sections and various distributions, see Ref.~\cite{Bardin:2008fn} for more
details. 

\section{Summary and outlook}

We have briefly presented the Monte Carlo event generator WINHAC for
precision theoretical prediction of the charged-current Drell--Yan process,
i.e. single $W$-boson production with leptonic decays in hadronic collisions.
It includes multiphoton radiation in the YFS exclusive exponentiation
scheme with the ISR-subtracted ${\cal O}(\alpha)$ EW corrections. For
QCD/QED initial-state parton shower as well as hadronisation, it is directly
interfaced with PYTHIA~6.4. It has been tested numerically and cross-checked
to the high precision with independent programs, in particular HORACE and SANC.
It has been used for developing new strategies of measuring the SM parameters
\cite{Krasny:2007cy, Fayette:2008wt,Fayette:2009zi,Upcoming_MW}
and testing the mechanism of the electroweak symmetry breaking 
\cite{Krasny:2005cb} at the LHC. 

The current version, {\tt 1.30}, of WINHAC is written in {\tt FORTRAN77},
however work is now in progress on re-writing it in {\tt C++}. 
A similar program, called ZINHAC~\cite{ZINHAC:MC}, 
for the neutral-current Drell--Yan process is being completed now 
-- in {\tt C++}.

\bibliographystyle{h-physrev3}

\end{document}